\documentclass[aps,graphicx]{revtex4}
\usepackage{graphicx}
\usepackage{bm}
\begin{document}

\title{Generation of concatenated Greenberger-Horne-Zeilinger-type entangled coherent state based on linear optics}

\author{Rui Guo,$^{1}$ Lan Zhou,$^{1,2}$ Shi-Pu Gu$^{3}$, Xing-Fu Wang$^{2}$ and Yu-Bo Sheng$^{1}$\footnote{shengyb@njupt.edu.cn}}
\address{$^1$ Key Lab of Broadband Wireless Communication and Sensor Network
 Technology,Nanjing University of Posts and Telecommunications, Ministry of
 Education, Nanjing, 210003,
 China\\
$^2$ College of Mathematics \& Physics, Nanjing University of Posts and Telecommunications, Nanjing,
210003, China\\
$^3$ College of Electronic Science and Engineering, Nanjing University of Posts and Telecommunications,
Nanjing 210003, China\\}

\begin{abstract}
The concatenated Greenberger-Horne-Zeilinger (C-GHZ)  state is a new type of multipartite entangled state, which has potential application in future quantum information. In this paper, we propose a protocol of constructing arbitrary C-GHZ entangled  state approximatively.  Different from the previous protocols, each logic is encoded in the coherent state. This protocol is based on the linear optics, which is feasible in experimental technology. This protocol may be useful in quantum information based on the C-GHZ state.
\end{abstract}\maketitle

\section{Introduction}
Quantum information, based on quantum mechanics theory, provides a great opportunity to develop computer and communication technology. Based on bipartite entanglement, many important quantum information protocols were developed,  such as quantum computation \cite{qc}, quantum teleportation \cite{teleportation1},  quantum key distribution \cite{QKD}, quantum secure direct communication \cite{QSDC1,QSDC2}, dense coding \cite{dense} and other significant quantum protocols \cite{other1,other2,other3,other4,other5}.
Besides the bipartite entanglement, multipartite entanglement also plays an important role in quantum information, such as controlled teleportation \cite{cteleportation}, quantum state sharing \cite{qss,qss1} and quantum secret sharing \cite{qss2}.
 Among  different types of multipartite  entanglement, the Greeberger-Horne-Zeilinger (GHZ) state has been widely used \cite{GHZ1}. However, due to the increase of the number of particle in noisy environment, the GHZ state will become fragile. Recently, Fr\"{o}w and D\"{u}r discussed a new type of multipartite entanglement, called   concatenated Greenberger-Horne-Zeilinger (C-GHZ) state\cite{CGHZ0}.
The C-GHZ state essentially is the logic-qubit entanglement and it  can be written as
\begin{eqnarray}
|\varphi\rangle_{N,m}&=&\frac{1}{\sqrt{2}}(|GHZ^{+}_{m}\rangle^{\otimes N}+|GHZ^{-}_{m}\rangle^{\otimes N}),\label{C-GHZ1}
\end{eqnarray}
with $|GHZ^{\pm}_{m}\rangle=\frac{1}{\sqrt{2}}(|0\rangle^{\otimes m}\pm|1\rangle^{\otimes m})$, where $N$ is the number of logical qubits. Each logic qubit is encoded by $m$ physical qubits.
Fr\"{o}w and D\"{u}r showed that the C-GHZ state not only has the similar feature of GHZ state, but also has a better robustness in  noisy environment \cite{CGHZ3}.  Subsequently, there are some interesting work based on the C-GHZ state, such as the generation of C-GHZ state with cross-Kerr nonlinearity \cite{construct1}, the Bell-state analysis for C-GHZ state \cite{cghzanalysis1,cghzanalysis2,cghzanalysis3,cghzanalysis4}, entanglement purification \cite{cghzpurification} and concentration \cite{cghzconcentration1,cghzconcentration2}. Based on the linear optics, Lu  \emph{et al.} designed the first experiment realization for C-GHZ state,  and showed that such state may has its potential application in future quantum information area \cite{construct2}.

On the other hand, in quantum information, there are two different approaches to encode the qubit.
The first  is the discrete variables (DV), such as the polarization states of photons \cite{polarization}, the spatial modes of the photons, and so on.  The second is the continuous variables (CV), such as the squeezed-state entanglement, Gaussian state  and coherent state \cite{squeezed0,squeezed1,squeezed2,squeezed3,squeezed4,Gaussian0,Gaussian1,Gaussian2,ECS0,ECS1,ECS2,ECS3,ECSadd2,ECS4,ECS5,ECS6,ECS7,ECSadd1,ECS8,ECSadd3,ECSadd4,ECSadd5}. For example, it is  shown that the squeezed state can be used in quantum teleportation, QKD and some other important quantum communication protocols \cite{squeezed0,squeezed1,squeezed2,squeezed3,squeezed4}.  In 2016, Faria described an optical approach for Gaussian state
 to verify bipartite entanglement. The most advantage is that this protocol does not destroy
both systems and their entanglement is proposed \cite{Gaussian1}.
The entangled coherent
state  (ECS) is another important type of CV entanglement, which will be detailed here.
In 2001, Wang \emph{et al.} described the quantum teleportation based on the ECS and  showed that the ECS can also be used in quantum communication \cite{ECS1}. In 2002, Jeong \emph{et al.} described an important protocol for  quantum computation with ECS \cite{ECS2}. Recently, the entanglement concentration for both Bell-type and W-type ECS were proposed \cite{ECS6,ECS7}.

 Though several quantum protocols for C-GHZ state were proposed, they all focus on the state encoded in DV, none protocol discuss the C-GHZ state in the framework of CV.
 In this paper,  we will propose the first approach to prepare the C-GHZ state encoded in coherent state approximatively, which is named C-GHZ-type ECS.  The qubit of C-GHZ-type ECS is encoded in coherent state, i. e.,   $|0\rangle \equiv|\alpha\rangle$ and $|1\rangle \equiv|-\alpha\rangle$, respectively.  Each logic qubit is an ECS. This protocol is based on  linear optics and may be feasible in future experiment.
This paper is organized as follows. In section 2, we briefly describe the protocol of constructing C-GHZ-type ECS with $N=m=2$ and $N=2$, $m=3$, respectively.
In section 3,  we extend this protocol to the case  C-GHZ-type ECS with arbitrary $N$ and $m$. In Section 4,  we  provide  discussion and  conclusion.

\section{Generation of C-GHZ-type ECS with $N=m=2$ and $N=2$, $m=3$}
In this section, we will propose a simple protocol to generate the C-GHZ-type ECS  approximatively. We first describe the approach to generate the C-GHZ-type ECS with
 $N=m=2$.  The C-GHZ-type ECS is
\begin{eqnarray}
|\psi\rangle_{2,2}&=&\frac{1}{\sqrt{2}}(|GHZ^{+}_{2}\rangle^{\otimes 2}+|GHZ^{-}_{2}\rangle^{\otimes 2})\nonumber\\\label{ECS1}
\end{eqnarray}
Here, state $|GHZ^{\pm}_{2}\rangle=[2(1\pm e^{-4|\alpha|^{2}})]^{-\frac{1}{2}}(|\alpha\rangle_{1}|\alpha\rangle_{2}\pm|-\alpha\rangle_{1}|-\alpha\rangle_{2})$.
States $|GHZ^{\pm}_{2}\rangle$  are the Bell-type ECS.
\begin{figure}[!h]
\begin{center}
\includegraphics[width=10cm,angle=0]{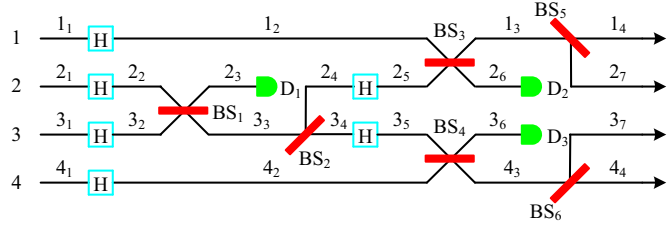}
\caption{Schematic diagram of generating the C-GHZ-type ECS with $N=m=2$. $H$ represents the Hadamard operation which makes $|\alpha\rangle\rightarrow\frac{N_{0}}{\sqrt{2}}(|\alpha\rangle+|-\alpha\rangle)$ and $|-\alpha\rangle\rightarrow\frac{N'_{0}}{\sqrt{2}}(|\alpha\rangle-|-\alpha\rangle)$. The BS is 50:50 beam splitter. }
\end{center}
\end{figure}

As shown in Fig.1, firstly, we prepare four coherent states $|\alpha\rangle$ in the spatial modes $1_{1}$, $2_{1}$, $3_{1}$ and $4_{1}$, respectively. The
four coherent states are $|\alpha\rangle_{1_{1}}$, $|\alpha\rangle_{2_{1}}$, $|\alpha\rangle_{3_{1}}$ and $|\alpha\rangle_{4_{1}}$, respectively. Subsequently,
we perform a  Hadamard operation on each single coherent state \cite{ECS8}. The Hadamard operation can make $|\alpha\rangle\rightarrow\frac{N_{0}}{\sqrt{2}}(|\alpha\rangle+|-\alpha\rangle)$ and $|-\alpha\rangle\rightarrow\frac{N'_{0}}{\sqrt{2}}(|\alpha\rangle-|-\alpha\rangle)$, respectively. Here, the normalized coefficient $N_{0}=(1+e^{-2|\alpha|^{2}})^{-\frac{1}{2}}$ and $N'_{0}=(1-e^{-2|\alpha|^{2}})^{-\frac{1}{2}}$. 
The state in spatial mode $2_{2}$ combined with the state in spatial mode $3_{2}$ can be written as
\begin{eqnarray}
|\psi\rangle_{1_{1}}&=&\frac{1}{\sqrt{2}}N_{0}(|\alpha\rangle_{2_{2}}+|-\alpha\rangle_{2_{2}})\otimes\frac{1}{\sqrt{2}}N_{0}(|\alpha\rangle_{3_{2}}+|-\alpha\rangle_{3_{2}})\nonumber\\
&=&\frac{1}{2}N_{0}^{2}(|\alpha\rangle_{2_{2}}|\alpha\rangle_{3_{2}}+|-\alpha\rangle_{2_{2}}|\alpha\rangle_{3_{2}}+|\alpha\rangle_{2_{2}}|-\alpha\rangle_{3_{2}}+|-\alpha\rangle_{2_{2}}|-\alpha\rangle_{3_{2}}).\label{ECS2}
\end{eqnarray}

Then, we let the photons in the spatial modes $2_{2}$ and $3_{2}$ pass through the 50:50 beam splitter (BS). The BS can transform  two different coherent states $|\alpha\rangle$ and $|\beta\rangle$ as
\begin{eqnarray}
|\alpha\rangle|\beta\rangle\rightarrow|\frac{\alpha+\beta}{\sqrt{2}}\rangle|\frac{\alpha-\beta}{\sqrt{2}}\rangle.\label{ECS3}
\end{eqnarray}
After passing through the BS$_{1}$, state $|\psi\rangle_{1_{1}}$ becomes
\begin{eqnarray}
|\psi\rangle_{1_{2}}&=&\frac{1}{2}N_{0}^{2}(|\sqrt{2}\alpha\rangle_{2_{3}}|0\rangle_{3_{3}}+|-\sqrt{2}\alpha\rangle_{2_{3}}|0\rangle_{3_{3}}+|0\rangle_{2_{3}}|\sqrt{2}\alpha\rangle_{3_{3}}+|0\rangle_{2_{3}}|-\sqrt{2}\alpha\rangle_{3_{3}}).\label{ECS2}
\end{eqnarray}

By choosing the cases that  the spatial mode $2_{3}$ has no photon, state $|\psi\rangle_{1_{2}}$ becomes
\begin{eqnarray}
|\psi\rangle_{1_{2}}\rightarrow|\psi\rangle_{1_{3}}&=&N_{1}(|0\rangle_{2_{3}}|\sqrt{2}\alpha\rangle_{3_{3}}+|0\rangle_{2_{3}}|-\sqrt{2}\alpha\rangle_{3_{3}}).
\end{eqnarray}
Here $N_{1}=[2(1+e^{-4|\alpha|^{2}})]^{-\frac{1}{2}}$.
Then we let the photons in the spatial mode $3_{3}$ pass through   BS$_{2}$, and then the state $|\psi\rangle_{1_{2}}$ will become
\begin{eqnarray}
|\psi\rangle_{1_{4}}&=&N_{1}(|\alpha\rangle_{2_{4}}|\alpha\rangle_{3_{4}}+|-\alpha\rangle_{2_{4}}|-\alpha\rangle_{3_{4}}).
\end{eqnarray}
Then, we perform a Hadamard operation on the spatial modes $2_{4}$ and $3_{4}$, respectively. The state $|\psi\rangle_{1_{4}}$ will evolve to
\begin{eqnarray}
|\psi\rangle_{1_{5}}&=&\frac{1}{2}N_{1}N_{0}^{2}[(|\alpha\rangle_{2_{5}}+|-\alpha\rangle_{2_{5}})(|\alpha\rangle_{3_{5}}+|-\alpha\rangle_{3_{5}})]
+\frac{1}{2}N_{1}N_{0}'^{2}[(|\alpha\rangle_{2_{5}}-|-\alpha\rangle_{2_{5}})(|\alpha\rangle_{3_{5}}-|-\alpha\rangle_{3_{5}})].\nonumber\\
\end{eqnarray}
The states in  spatial modes $1_{2}$ and $4_{2}$ combined with $|\psi\rangle_{1_{5}}$ can be rewritten as
\begin{eqnarray}
|\psi\rangle_{1_{6}}&=&|\psi\rangle_{1_{5}}\otimes\frac{1}{\sqrt{2}}N_{0}(|\alpha\rangle_{1_{2}}+|-\alpha\rangle_{1_{2}})\otimes\frac{1}{\sqrt{2}}N_{0}(|\alpha\rangle_{4_{2}}+|-\alpha\rangle_{4_{2}})\nonumber\\
&=&\frac{1}{4}N_{0}^{4}N_{1}[(|\alpha\rangle_{1_{2}}+|-\alpha\rangle_{1_{2}})(|\alpha\rangle_{2_{5}}+|-\alpha\rangle_{2_{5}})(|\alpha\rangle_{3_{5}}+|-\alpha\rangle_{3_{5}})(|\alpha\rangle_{4_{2}}+|-\alpha\rangle_{4_{2}})]\nonumber\\
&+&\frac{1}{4}N_{0}^{2}N_{0}'^{2}N_{1}[(|\alpha\rangle_{1_{2}}+|-\alpha\rangle_{1_{2}})(|\alpha\rangle_{2_{5}}-|-\alpha\rangle_{2_{5}})(|\alpha\rangle_{3_{5}}-|-\alpha\rangle_{3_{5}})(|\alpha\rangle_{4_{2}}+|-\alpha\rangle_{4_{2}})].\nonumber\\
\end{eqnarray}
 We let the photons in the spatial modes $1_{2}$ and $2_{5}$ pass through BS$_{3}$, and the photons in the spatial modes $3_{5}$ and $4_{2}$ pass through BS$_{4}$, respectively.  The state $|\psi\rangle_{1_{6}}$ can be written as
\begin{eqnarray}
|\psi\rangle_{1_{7}}&=&\frac{1}{4}N_{0}^{4}N_{1}[(|\sqrt{2}\alpha\rangle_{1_{3}}|0\rangle_{2_{6}}+|-\sqrt{2}\alpha\rangle_{1_{3}}|0\rangle_{2_{6}}+|0\rangle_{1_{3}}|\sqrt{2}\alpha\rangle_{2_{6}}+|0\rangle_{1_{3}}|-\sqrt{2}\alpha\rangle_{2_{6}})\nonumber\\
&&\otimes(|\sqrt{2}\alpha\rangle_{3_{6}}|0\rangle_{4_{3}}+|-\sqrt{2}\alpha\rangle_{3_{6}}|0\rangle_{4_{3}}+|0\rangle_{3_{6}}|\sqrt{2}\alpha\rangle_{4_{3}}+|0\rangle_{3_{6}}|-\sqrt{2}\alpha\rangle_{4_{3}})]\nonumber\\
&&+\frac{1}{4}N_{0}^{2}N_{0}'^{2}N_{1}[(|\sqrt{2}\alpha\rangle_{1_{3}}|0\rangle_{2_{6}}-|-\sqrt{2}\alpha\rangle_{1_{3}}|0\rangle_{2_{6}}-|0\rangle_{1_{3}}|\sqrt{2}\alpha\rangle_{2_{6}}+|0\rangle_{1_{3}}|-\sqrt{2}\alpha\rangle_{2_{6}})\nonumber\\
&&\otimes(|\sqrt{2}\alpha\rangle_{3_{6}}|0\rangle_{4_{3}}-|-\sqrt{2}\alpha\rangle_{3_{6}}|0\rangle_{4_{3}}+|0\rangle_{3_{6}}|\sqrt{2}\alpha\rangle_{4_{3}}-|0\rangle_{3_{6}}|-\sqrt{2}\alpha\rangle_{4_{3}})].
\end{eqnarray}
By selecting the case that both spatial modes $2_{6}$ and $3_{6}$ have no photon,  state $|\psi\rangle_{1_{7}}$ will collapse to
\begin{eqnarray}
|\psi\rangle_{1_{8}}&=&\frac{1}{4}N_{0}^{4}N_{1}[(|\sqrt{2}\alpha\rangle_{1_{3}}|0\rangle_{2_{6}}+|-\sqrt{2}\alpha\rangle_{1_{3}}|0\rangle_{2_{6}})(|0\rangle_{3_{6}}|\sqrt{2}\alpha\rangle_{4_{3}}+|0\rangle_{3_{6}}|-\sqrt{2}\alpha\rangle_{4_{3}})]\nonumber\\
&&+\frac{1}{4}N_{0}^{2}N_{0}'^{2}N_{1}[(|\sqrt{2}\alpha\rangle_{1_{3}}|0\rangle_{2_{6}}-|-\sqrt{2}\alpha\rangle_{1_{3}}|0\rangle_{2_{6}})(|0\rangle_{3_{6}}|\sqrt{2}\alpha\rangle_{4_{3}}-|0\rangle_{3_{6}}|-\sqrt{2}\alpha\rangle_{4_{3}})].\nonumber\\
\end{eqnarray}

Finally, we let the photons in spatial mode $1_{3}$ pass through BS$_{5}$, and let the photons in spatial mode $4_{3}$ pass through BS$_{6}$. We can ultimately obtain the C-GHZ-type ECS in the spatial modes $1_{4}$, $2_{7}$, $3_{7}$ and $4_{4}$ as
\begin{eqnarray}
|\psi\rangle_{1_{9}}&=&\frac{1}{4}N_{0}^{4}N_{1}(|\alpha\rangle_{1_{4}}|\alpha\rangle_{2_{7}}+|-\alpha\rangle_{1_{4}}|-\alpha\rangle_{2_{7}})
(|\alpha\rangle_{3_{7}}|\alpha\rangle_{4_{4}}+|-\alpha\rangle_{3_{7}}|-\alpha\rangle_{4_{4}})\nonumber\\
&+&\frac{1}{4}N_{0}^{2}N_{0}'^{2}N_{1}(|\alpha\rangle_{1_{4}}|\alpha\rangle_{2_{7}}+|-\alpha\rangle_{1_{4}}|-\alpha\rangle_{2_{7}})
(|\alpha\rangle_{3_{7}}|\alpha\rangle_{4_{4}}-|-\alpha\rangle_{3_{7}}|-\alpha\rangle_{4_{4}})\nonumber\\
&\approx&\frac{1}{\sqrt{2}}(|GHZ^{+}_{2}\rangle^{\otimes 2}+|GHZ^{-}_{2}\rangle^{\otimes 2}).\label{ECS2-2}
\end{eqnarray}
From Eq. (\ref{ECS2-2}), in order to obtain the state in Eq. (\ref{ECS1}), if $\alpha$ is large enough, which makes $e^{-2|\alpha|^{2}}\rightarrow 0$.

This approach can be extend to the case of generating  the C-GHZ-type ECS with  $N=2$, and $m=3$ of the form
\begin{eqnarray}
|\psi\rangle_{2,3}&=&\frac{1}{\sqrt{2}}(|GHZ^{+}_{3}\rangle^{\otimes 2}+|GHZ^{-}_{3}\rangle^{\otimes 2}).\label{cghz23}
\end{eqnarray}
Here, $|GHZ^{\pm}_{3}\rangle=[2(1\pm e^{-6|\alpha|^{2}})]^{-\frac{1}{2}}(|\alpha\rangle_{1}|\alpha\rangle_{2}|\alpha\rangle_{3}\pm|-\alpha\rangle_{1}|-\alpha\rangle_{2}|-\alpha\rangle_{3})$.
 From Eq. (\ref{cghz23}), in each logic qubit, it is a GHZ-type ECS.
\begin{figure}[!h]
\begin{center}
\includegraphics[width=12cm,angle=0]{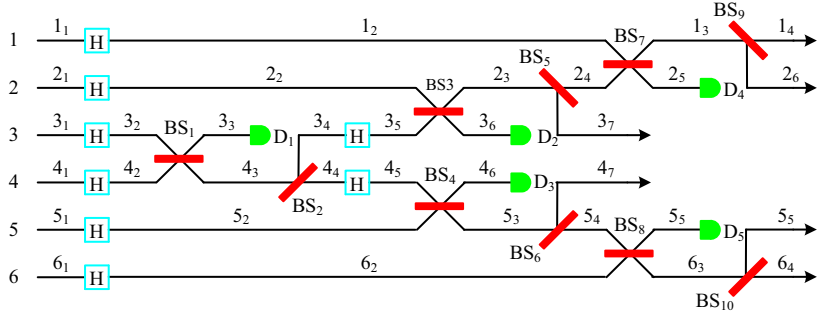}
\caption{Schematic diagram of generating the C-GHZ-type ECS with $m=3$ and $N=2$.}
\end{center}
\end{figure}
As shown in Fig.2, we first prepare six coherent states $|\alpha\rangle$ in spatial modes $1_{1}$, $2_{1}$, $3_{1}$, $4_{1}$, $5_{1}$ and $6_{1}$. With the same principle shown in Fig. 1, we first generate the state
\begin{eqnarray}
|\psi\rangle_{2_{1}}&\rightarrow&\frac{1}{\sqrt{2}}N_{1}^{2}[(|\alpha\rangle_{2_{4}}|\alpha\rangle_{3_{7}}+|-\alpha\rangle_{2_{4}}|-\alpha\rangle_{3_{7}})(|\alpha\rangle_{4_{7}}|\alpha\rangle_{5_{4}}+|-\alpha\rangle_{4_{7}}|-\alpha\rangle_{5_{4}})]\nonumber\\
&&+\frac{1}{\sqrt{2}}N_{1}'^{2}[(|\alpha\rangle_{2_{4}}|\alpha\rangle_{3_{7}}-|-\alpha\rangle_{2_{4}}|-\alpha\rangle_{3_{7}})(|\alpha\rangle_{4_{7}}|\alpha\rangle_{5_{4}}-|-\alpha\rangle_{4_{7}}|-\alpha\rangle_{5_{4}})].\label{ECS7}
\end{eqnarray}
in the spatial modes $2_{4}$, $3_{7}$, $4_{7}$ and $5_{4}$, which has the same form of the state in Eq. (\ref{ECS1}). Here $N'_{1}=[2(1- e^{-4|\alpha|^{2}})]^{-\frac{1}{2}}$.

In next step, we let the  photons in spatial modes $1_{2}$ and $2_{4}$ pass through BS$_{7}$, and let the photons in spatial modes $5_{4}$ and $6_{2}$ pass through BS$_{8}$, respectively. The state $|\psi\rangle_{2_{1}}$ combined with the state in  spatial modes $1_{2}$ and $6_{2}$ becomes
\begin{eqnarray}
|\psi\rangle_{2_{2}}&=&|\psi\rangle_{2_{1}}\otimes\frac{1}{\sqrt{2}}N_{0}(|\alpha\rangle_{1_{2}}+|-\alpha\rangle_{1_{2}})\otimes\frac{1}{\sqrt{2}}N_{0}(|\alpha\rangle_{6_{2}}+|-\alpha\rangle_{6_{2}})\nonumber\\
&\rightarrow&\frac{1}{2\sqrt{2}}N_{0}^{2}N_{1}^{2}[(|\sqrt{2}\alpha\rangle_{1_{3}}|0\rangle_{2_{5}}|\alpha\rangle_{3_{7}}+|0\rangle_{1_{3}}|-\sqrt{2}\alpha\rangle_{2_{5}}|\alpha\rangle_{3_{7}}+|0\rangle_{1_{3}}|\sqrt{2}\alpha\rangle_{2_{5}}|-\alpha\rangle_{3_{7}}\nonumber\\
&+&|-\sqrt{2}\alpha\rangle_{1_{3}}|0\rangle_{2_{5}}|-\alpha\rangle_{3_{7}})(|\alpha\rangle_{4_{7}}|\sqrt{2}\alpha\rangle_{5_{5}}|0\rangle_{6_{3}}+|-\alpha\rangle_{4_{7}}0\rangle_{5_{5}}|-\sqrt{2}\alpha\rangle_{6_{3}}\nonumber\\
&+&|\alpha\rangle_{4_{7}}|0\rangle_{5_{7}}|\sqrt{2}\alpha\rangle_{6_{3}}+|-\alpha\rangle_{4_{7}}|-\sqrt{2}\alpha\rangle_{5_{5}}|0\rangle_{6_{3}})]\nonumber\\
&+&\frac{1}{2\sqrt{2}}N_{0}^{2}N_{1}'^{2}[(|\sqrt{2}\alpha\rangle_{1_{3}}|0\rangle_{2_{5}}|\alpha\rangle_{3_{7}}+|0\rangle_{1_{3}}|-\sqrt{2}\alpha\rangle_{2_{5}}|\alpha\rangle_{3_{7}}-|0\rangle_{1_{3}}|\sqrt{2}\alpha\rangle_{2_{5}}|-\alpha\rangle_{3_{7}}\nonumber\\
&-&|-\sqrt{2}\alpha\rangle_{1}|0\rangle_{2}|-\alpha\rangle_{3})(|\alpha\rangle_{4_{7}}|\sqrt{2}\alpha\rangle_{5_{5}}|0\rangle_{6_{3}}-|-\alpha\rangle_{4_{7}}|0\rangle_{5_{5}}|-\sqrt{2}\alpha\rangle_{6_{3}}\nonumber\\
&+&|\alpha\rangle_{4_{7}}|0\rangle_{5_{5}}|\sqrt{2}\alpha\rangle_{6_{3}}-|-\alpha\rangle_{4_{7}}|-\sqrt{2}\alpha\rangle_{5_{5}}|0\rangle_{6_{3}})].\label{ECS10}
\end{eqnarray}

Finally,  we choose the cases that both spatial modes $2_{5}$ and $5_{5}$ have no photon, and let the state photons in spatial mode $1_{3}$ pass through BS$_{9}$, and  the photons in spatial mode $6_{3}$ pass through BS$_{10}$, respectively. The state $|\psi\rangle_{2_{2}}$ will ultimately collapse to state
\begin{eqnarray}
|\psi\rangle_{2_{3}}&\rightarrow&\frac{1}{2\sqrt{2}}N_{0}^{2}N_{1}^{2}[(|\sqrt{2}\alpha\rangle_{1_{3}}|0\rangle_{2_{5}}|\alpha\rangle_{3_{7}}+|-\sqrt{2}\alpha\rangle_{1_{3}}|0\rangle_{2_{5}}|-\alpha\rangle_{3_{7}})\nonumber\\
&&\otimes(|-\alpha\rangle_{4_{7}}|0\rangle_{5_{5}}|-\sqrt{2}\alpha\rangle_{6_{3}}+|\alpha\rangle_{4_{7}}|0\rangle_{5_{5}}|\sqrt{2}\alpha\rangle_{6_{3}})\nonumber\\
&&+\frac{1}{2\sqrt{2}}N_{0}^{2}N_{1}'^{2}[(|\sqrt{2}\alpha\rangle_{1_{3}}|0\rangle_{2_{5}}|\alpha\rangle_{3_{7}}-|-\sqrt{2}\alpha\rangle_{1_{3}}|0\rangle_{2_{5}}|-\alpha\rangle_{3_{7}})]\nonumber\\
&&\otimes(|\alpha\rangle_{4_{7}}|0\rangle_{5_{5}}|\sqrt{2}\alpha\rangle_{6_{3}}-|-\alpha\rangle_{4_{7}}|0\rangle_{5_{5}}|-\sqrt{2}\alpha\rangle_{6_{3}})]\nonumber\\
&\rightarrow&\frac{1}{2\sqrt{2}}N_{0}^{2}N_{1}^{2}[(|\alpha\rangle_{1_{4}}|\alpha\rangle_{2_{6}}|\alpha\rangle_{3_{7}}+|-\alpha\rangle_{1_{4}}|-\alpha\rangle_{2_{6}}|-\alpha\rangle_{3_{7}})\nonumber\\
&&\otimes(|\alpha\rangle_{4_{7}}|\alpha\rangle_{5_{5}}|\alpha\rangle_{6_{4}}+|-\alpha\rangle_{4_{7}}|-\alpha\rangle_{5_{5}}|-\alpha\rangle_{6_{4}})]\nonumber\\
&&+\frac{1}{2\sqrt{2}}N_{0}^{2}N_{1}'^{2}[(|\alpha\rangle_{1_{4}}|\alpha\rangle_{2_{6}}|\alpha\rangle_{3_{7}}-|-\alpha\rangle_{1_{4}}|-\alpha\rangle_{2_{6}}|-\alpha\rangle_{3_{7}})\nonumber\\
&&\otimes(|\alpha\rangle_{4_{7}}|\alpha\rangle_{5_{5}}|\alpha\rangle_{6_{4}}-|-\alpha\rangle_{4_{7}}|-\alpha\rangle_{5_{5}}|-\alpha\rangle_{6_{4}})].\label{ECS12}
\end{eqnarray}
State in Eq. (\ref{ECS12}) is the obtained state as shown in Eq. (\ref{cghz23}) if $e^{-2|\alpha|^{2}}\rightarrow 0$.

\section{Generation of C-GHZ-type ECS with arbitrary $N$ and $m$}
It is straightforward to extend this protocol to the case of    generation of the C-GHZ-type ECS with arbitrary $N$ and $m$.
\begin{figure}[!h]
\begin{center}
\includegraphics[width=11cm,angle=0]{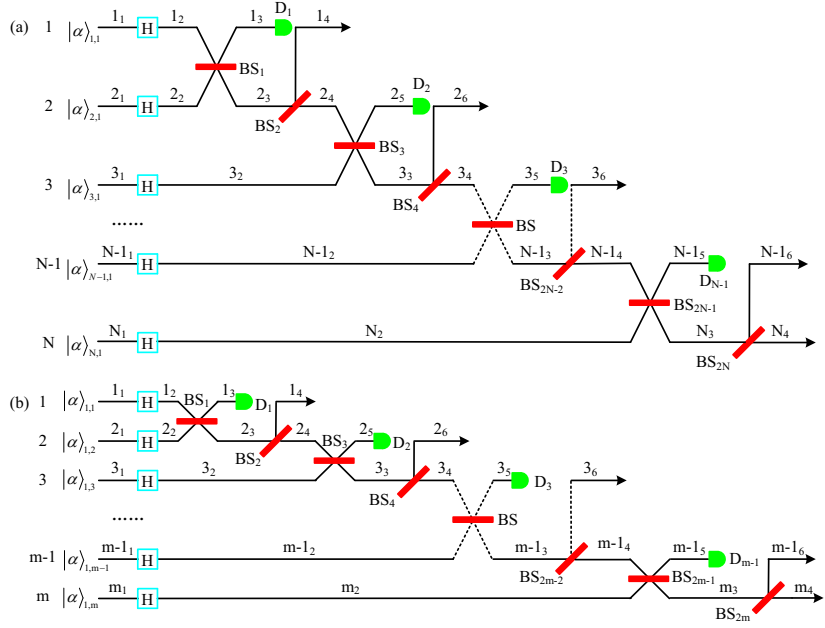}
\caption{Schematic diagram of generating the C-GHZ-type ECS with arbitrary $N$ and $m$. (a) Schematic diagram shows the generation of GHZ-type ECS. (b) Schematic diagram shows the generation of logical bits for the first coherent state in the GHZ-type ECS.
 }
\end{center}
\end{figure}
Generally, the whole protocol can be divided into two steps.
In the first step, we generate a GHZ-type ECS with $N$ coherent states. As shown in Fig. 3 (a), we prepare $N$ coherent states in the spatial modes
$1_{1}$, $2_{1}$, $\cdots$, $N_{1}$. Similar to the previous section, we first generate Bell-type ECS in the spatial modes $1_{4}$ and $2_{4}$ of the form
\begin{eqnarray}
|\Psi\rangle_{1_{1}}&=&N_{1}(|\alpha\rangle_{1_{4}}|\alpha\rangle_{2_{4}}+|-\alpha\rangle_{1_{4}}|-\alpha\rangle_{2_{4}}),
\end{eqnarray}
under the condition that single-photon detector D$_{1}$ detect no photon.
Then, we combine the state $|\Psi\rangle_{1_{1}}$ with the coherent state in spatial mode $3_{2}$.  The whole sytem can be written as
\begin{eqnarray}
|\Psi\rangle_{1_{2}}&=&|\Psi\rangle_{1_{1}}\otimes\frac{1}{\sqrt{2}}N_{0}(|\alpha\rangle_{3_{2}}+|-\alpha\rangle_{3_{2}})\nonumber\\
&=&\frac{1}{\sqrt{2}}N_{0}N_{1}(|\alpha\rangle_{1_{4}}|\alpha\rangle_{2_{4}}|\alpha\rangle_{3_{2}}+|-\alpha\rangle_{1_{4}}|-\alpha\rangle_{2_{4}}|\alpha\rangle_{3_{2}}\nonumber\\
&&+|\alpha\rangle_{1_{4}}|\alpha\rangle_{2_{4}}|-\alpha\rangle_{3_{2}}+|-\alpha\rangle_{1_{4}}|-\alpha\rangle_{2_{4},1}|-\alpha\rangle_{3_{2}}).
\end{eqnarray}

 We let the photons in spatial modes $2_{4}$ and $3_{2}$ pass through BS $_{3}$ and choosing  the case that the spatial mode $2_{5}$ has no photon. The state can be written as
\begin{eqnarray}
|\Psi\rangle_{1_{3}}&=&\frac{1}{\sqrt{2}}N_{0}N_{1}(|\alpha\rangle_{1_{4}}|\sqrt{2}\alpha\rangle_{2_{5}}|0\rangle_{3_{3}}+|-\alpha\rangle_{1_{4}}|0\rangle_{2_{5}}|-\sqrt{2}\alpha\rangle_{3_{3}}\nonumber\\
&+&|\alpha\rangle_{1_{4}}|0\rangle_{2_{5}}|\sqrt{2}\alpha\rangle_{3_{3}}+|-\alpha\rangle_{1_{4}}|-\sqrt{2}\alpha\rangle_{2_{5},1}|0\rangle_{3_{3}})\nonumber\\
&\rightarrow&\frac{1}{2}N_{0}N_{1}(|-\alpha\rangle_{1_{4}}|0\rangle_{2_{5}}|-\sqrt{2}\alpha\rangle_{3_{3}}+|\alpha\rangle_{1_{4}}|0\rangle_{2_{5}}|\sqrt{2}\alpha\rangle_{3_{3}}).
\end{eqnarray}

Let the photons in spatial mode $3_{3}$ pass through BS$_{4}$, and then the state will be described as
\begin{eqnarray}
|\Psi\rangle_{1_{3}}&=&N_{2}(|\alpha\rangle_{1_{4}}|\alpha\rangle_{2_{6}}|\alpha\rangle_{3_{4}}+|-\alpha\rangle_{1_{4}}|-\alpha\rangle_{2_{6}}|-\alpha\rangle_{3_{4}}).
\end{eqnarray}
Here $N_{2}=[2(1\pm e^{-6|\alpha|^{2}})]^{-\frac{1}{2}}$.
Following the same principle, by add another coherent states in spatial modes $3_{1}$, $4_{1}$, $\cdots$, $N_{1}$, we can get the GHZ-type ECS as
\begin{eqnarray}
|\Psi\rangle&=&N_{N}(|\alpha\rangle_{1_{4}}|\alpha\rangle_{2_{6}}|\alpha\rangle_{3_{6}}\cdot\cdot\cdot|\alpha\rangle_{N-2_{6}}|\alpha\rangle_{N-1_{6}}|\alpha\rangle_{N_{4}}\nonumber\\
&&+|-\alpha\rangle_{1_{4}}|-\alpha\rangle_{2_{6}}|-\alpha\rangle_{3_{6}}\cdot\cdot\cdot|-\alpha\rangle_{N-2_{6}}|-\alpha\rangle_{N-1_{6}}|-\alpha\rangle_{N_{4}})\nonumber\\
&=&N_{N}(|\alpha\rangle^{\otimes N}+|-\alpha\rangle^{\otimes N}).\label{GHZ}
\end{eqnarray}
Here $N_{N}=[2(1\pm e^{-2N|\alpha|^{2}})]^{-\frac{1}{2}}$.
In the second step, we  will generate the GHZ-type ECS on each qubit.
We rewrite the state $|\Psi\rangle$ as
\begin{eqnarray}
|\Psi\rangle_{1}&=&N_{N}(|\alpha\rangle_{1,1}|\alpha\rangle^{\otimes N-1}+|-\alpha\rangle_{1,1}|-\alpha\rangle^{\otimes N-1}).\label{GHZ1}
\end{eqnarray}
Here the first $"1"$ in subscript $"1,1"$ means the first logic qubit. The second $"1"$ means the first physical qubit.
As shown in Fig. 3 (b), we first perform the Hadamard operation on the qubit $1,1$ and make the state $|\Psi\rangle_{1}$ becomes
\begin{eqnarray}
|\Psi\rangle_{1}&\rightarrow&|\Psi\rangle_{2}=\frac{1}{\sqrt{2}}N_{N}N_{0}(|\alpha\rangle_{1,1_{2}}+|-\alpha\rangle_{1,1_{2}})|\alpha\rangle^{\otimes N-1}\nonumber\\
&+&\frac{1}{\sqrt{2}}N_{N}N'_{0}(|\alpha\rangle_{1,1_{2}}-|-\alpha\rangle_{1,1_{2}})|-\alpha\rangle^{\otimes N-1}.\label{GHZ3}
\end{eqnarray}
The state in Eq.(\ref{GHZ3}) combined with the single coherent state in the spatial mode $2_{2}$ can be written as
\begin{eqnarray}
|\Phi\rangle_{1}&=&|\Psi\rangle_{2}\otimes\frac{1}{\sqrt{2}}N_{0}(|\alpha\rangle_{1,2_{2}}+|-\alpha\rangle_{1,2_{2}}).
\end{eqnarray}
Then, we let the photons in spatial modes $1_{2}$ and $2_{2}$ pass through BS$_{1}$. The state will be
\begin{eqnarray}
|\Phi\rangle_{2}&=&\frac{1}{2}N^{2}_{0}N_{N}[(|\sqrt{2}\alpha\rangle_{1,1_{3}}|0\rangle_{1,2_{3}}+|0\rangle_{1,1_{3}}|\sqrt{2\alpha}\rangle_{1,2_{3}}\nonumber\\
&&+|0\rangle_{1,1_{3}}|-\sqrt{2}\alpha\rangle_{1,2_{3}}+|-\sqrt{2}\alpha\rangle_{1,1_{3}}|0\rangle_{1,2_{3}})|\alpha\rangle^{\otimes N-1}]\nonumber\\
&&+\frac{1}{2}N_{N}N_{0}N'_{0}[(|\sqrt{2}\alpha\rangle_{1,1_{3}}|0\rangle_{1,2_{3}}+|0\rangle_{1,1_{3}}|\sqrt{2\alpha}\rangle_{1,2_{3}}\nonumber\\
&&-|0\rangle_{1,1_{3}}|-\sqrt{2}\alpha\rangle_{1,2_{3}}-|-\sqrt{2}\alpha\rangle_{1,1_{3}}|0\rangle_{1,2_{3}})|-\alpha\rangle^{\otimes N-1}].
\end{eqnarray}
By choosing the cases where the spatial mode $1_{3}$ has no photon and making the photons in spatial mode $2_{2}$ pass through BS$_{2}$, we will obtain the state \begin{eqnarray}
|\Phi\rangle_{3}&=&\frac{1}{2}N^{2}_{0}N_{N}(|\alpha\rangle_{1,1_{4}}|\alpha\rangle_{1,2_{4}}+|-\alpha\rangle_{1,1_{4}}|-\alpha\rangle_{1,2_{4}})|\alpha\rangle^{\otimes N-1}\nonumber\\
&&+\frac{1}{2}N_{N}N_{0}N'_{0}(|\alpha\rangle_{1,1_{4}}|\alpha\rangle_{1,2_{4}}-|-\alpha\rangle_{1,1_{4}}|-\alpha\rangle_{1,2_{4}})|-\alpha\rangle^{\otimes N-1}.
\end{eqnarray}
Following the same principle, by adding another coherent states in spatial modes $3_{1}$, $4_{1}$, $\cdots$, $m_{1}$, we can obtain the state
\begin{eqnarray}
|\Phi\rangle_{1,m}&=&\frac{1}{2}N_{0}^{m-1}N_{N}(|\alpha\rangle_{1,1_{4}}|\alpha\rangle_{1,2_{6}}|\alpha\rangle_{1,3_{6}}\cdot\cdot\cdot|\alpha\rangle_{1,m-1_{6}}|\alpha\rangle_{1,m_{4}}\nonumber\\
&&+|-\alpha\rangle_{1,1_{4}}|-\alpha\rangle_{1,2_{6}}|-\alpha\rangle_{1,3_{6}}\cdot\cdot\cdot|-\alpha\rangle_{1,m-1_{6}}|-\alpha\rangle_{1,m_{4}})|\alpha\rangle^{\otimes N-1}\nonumber\\
&&+\frac{1}{2}N_{0}N_{N}N'^{m-2}_{0}(|\alpha\rangle_{1,1_{4}}|\alpha\rangle_{1,2_{6}}|\alpha\rangle_{1,3_{6}}\cdot\cdot\cdot|\alpha\rangle_{1,m-1_{6}}|\alpha\rangle_{1,m_{4}}\nonumber\\
&&-|-\alpha\rangle_{1,1_{4}}|-\alpha\rangle_{1,2_{6}}|-\alpha\rangle_{1,3_{6}}\cdot\cdot\cdot|-\alpha\rangle_{1,m-1_{6}}|-\alpha\rangle_{1,m_{4}})|-\alpha\rangle^{\otimes N-1}\nonumber\\
&=&\frac{1}{2}N_{0}^{m-1}N_{N}(|\alpha\rangle_{1}^{\otimes m}+|-\alpha\rangle_{1}^{\otimes m})|\alpha\rangle^{\otimes N-1}\nonumber\\
&&+\frac{1}{2}N_{0}N_{N}N'^{m-2}_{0}(|\alpha\rangle_{1}^{\otimes m}-|-\alpha\rangle_{1}^{\otimes m})|-\alpha\rangle^{\otimes N-1}.
\end{eqnarray}

In the same way, by adding $m-1$ coherent state in each logic qubit, we can finally obtain the  C-GHZ-type ECS
\begin{eqnarray}
|\Phi\rangle&=&(N^{m-2}_{0})^{\otimes N}((|\alpha\rangle^{\otimes m}+|-\alpha\rangle^{\otimes m})^{\otimes N}+(N'^{m-2}_{0})^{N}(|\alpha\rangle^{\otimes m}-|-\alpha\rangle^{\otimes m})^{\otimes \otimes N})\nonumber\\
&\approx&\frac{1}{\sqrt{2}}(|GHZ^{+}_{m}\rangle^{\otimes N}+|GHZ^{-}_{m}\rangle^{\otimes N}).
\end{eqnarray}

\section{Conclusion}
We have designed the protocol to construct the arbitrary C-GHZ-type ECS from single coherent states. Generally speaking, the protocol can be divided into two steps. In the first step, we prepare the N-GHZ-type ECS.
Second, we prepare  arbitrary  C-GHZ-type ECS from N-GHZ-type ECS. We should point out that this protocol cannot obtain the exact the C-GHZ-type ECS. The main reason is that the $\langle-\alpha|\alpha\rangle=e^{-2|\alpha|^{2}}\neq 0$. As shown in Eq. (\ref{ECS2-2}),  we can obtain the state in Eq. (\ref{ECS1}) under the condition that $e^{-2|\alpha|^{2}}\rightarrow 0$, if $\alpha$ is large enough. 

\begin{figure}[!h]
\begin{center}
\includegraphics[width=9cm,angle=0]{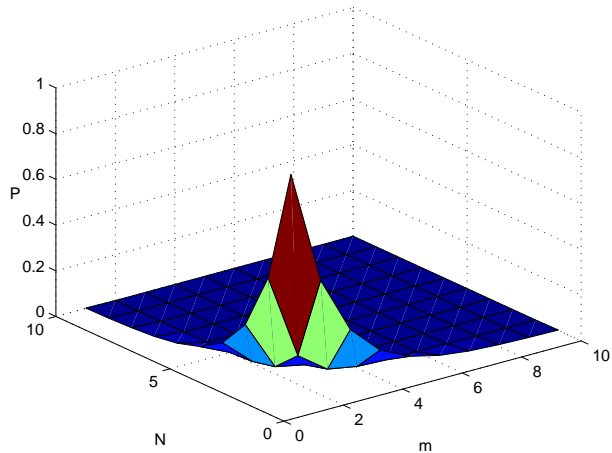}
\caption{Schematic diagram showing the success probability $P$ of the protocol. The $\alpha$ is the amplitude of the coherent state. Here, we let the $\alpha$=10. }
\end{center}
\end{figure}
The total success probability $P$ of the protocol can be written as
\begin{eqnarray}
P&=&\frac{1}{2^{Nm-1}}.
\end{eqnarray}
From Fig. 4, $P$ decreases with both $m$ and $N$. Here we choose $\alpha=10$.
In this protocol, the success case is the single-photon detector does not register any photon. Such selection condition provides us a good advantage to realize
this protocol. In conventional quantum information protocols based on linear optics, the success condition is that the single-photon detector register one and only one photon. However, once the photon is detected, it is also destroyed and  cannot be used in the future quantum information processing.
Interestingly, by selecting the case that the single-photon detector does not register photon, the state can be remained.
Certainly, we should point out that the coherent state $|\alpha\rangle$ also has the probability to make the single-photon detector do not register the photon. It will induce the error.  Fortunately, the error probability is too small and can be ignored,  if $\alpha$ is  large. For example, when $\alpha=2$, the error probability $|\langle0|\alpha\rangle|^{2}$ is $1.1\times10^{-7}$.

In summary, we have presented a protocol for preparing  the C-GHZ-type ECS.  We first describe the protocol in the case of $N=m=2$ and $N=2$, $m=3$ respectively. Then we describe the protocol in the case of $N=m=3$ and finally extend the protocol to the arbitrary number of $N$ and $m$.
 This protocol may be be useful in the future of long-distance quantum communications based on C-GHZ state.

\section{ACKNOWLEDGMENTS}
This work was supported by the National Natural Science Foundation
of China under Grant  Nos. 11474168 and 61401222, the Natural Science Foundation of Jiangsu province under Grant No. BK20151502, the Qing Lan Project in Jiangsu Province,  and a Project
Funded by the Priority Academic Program Development of Jiangsu Higher Education Institutions.


\begin{thebibliography}{26}

\bibitem{qc} Divincenzo D.P.: Quantum Computation. Science  \textbf{270}, 255  (1995)

\bibitem{teleportation1}Bennett, C.H., Brassard G., Crepeau C., Jozsa R., Peres A. and Wootters W. K.: Teleporting an unknown quantum state via dual classical and
einstein-podolsky-rosen channels.
Phys. Rev. Lett. \textbf{70}, 1895 (1993)

\bibitem{QKD} Ekert A.K.: Quantum cryptography based on Bell Theorem.  Phys. Rev. Lett. \textbf{67}, 661 (1991)


\bibitem{QSDC1}Long, G.L., Liu, X.S.: Theoretically efficient high-capacity quantum-key-distribution scheme.
Phys. Rev. A \textbf{65}, 032302 (2002)

\bibitem{QSDC2} Deng, F.G., Long, G.L., Liu, X.S.: Two-step quantum direct communication protocol using the
einstein-podolsky-rosen pair block.
Phys. Rev. A \textbf{68}, 042317 (2003)


\bibitem{dense} Bennett C.H., Brassard G. and Mermin N.D.: Quantum cryptography without Bell Theorem. Phys. Rev. Lett. \textbf{68}, 557 (1992)

\bibitem{other1}Zheng, C., Long, G.F.: Quantum secure direct dialogue using Einstein-Podolsky-Rosen pairs. Sci.
Chin. Phys. Mech. Astron. \textbf{57}, 1238-1243 (2014)

\bibitem{other2}Cao, D.Y., Liu, B.H.,Wang, Z., Huang, Y.F., Li, C.F., Guo, G.C.: Multiuser-to-multiuser entanglement
distribution based on 1550 nm polarization-entangled photons. Sci. Bull. \textbf{60}, 1128-1132 (2015)

\bibitem{other3} Su, X.L., Jia, X.J., Xie, C.D., Peng, K.C.: Preparation of multipartite entangled states used for quantum
information networks. Sci. Chin. Phys. Mech. Astron. \textbf{57}, 1210-1217 (2014)

\bibitem{other4}Zhang, C., Li, C.F., Guo, G.C.: Experimental demonstration of photonic quantum ratchet. Sci. Bull.
\textbf{60}, 249-255 (2015)

\bibitem{other5}Zou, X.F., Qiu, D.W.: Three-step semiquantum secure direct communication protocol. Sci. Chin. Phys.
Mech. Astro. \textbf{57}, 1696-1702 (2014)

\bibitem{cteleportation}Deng, F. G., Li, C. Y., Li, Y. S., Zhou, H. Y., Wang, Y.:Symmetric multiparty-controlled teleportation of an arbitrary two-particle entanglement. Phys. Rev. A \textbf{72}, 022338 (2005)

\bibitem{qss}Cleve, R., Gottesman, D.,  Lo, H. K.:How to share a quantum secret.  Phys. Rev. Lett. \textbf{83},
648 (1999)

\bibitem{qss1}Deng, F. G.,  Li, X. H., Li, C. Y. Zhou,  P.  Zhou, H. Y.:Multiparty quantum-state sharing of an arbitrary two-particle state with Einstein-Podolsky-Rosen pairs.  Phys. Rev. A \textbf{72}, 044301 (2005)

\bibitem{qss2}Hillery, M., Bu\u{z}ek,  V.,  Berthiaume A.:Quantum secret sharing. Phys. Rev. A \textbf{59},
1829 (1999)

\bibitem{GHZ1} Pan, J. W., Chen, Z. B., Lu, C. Y., Weinfurter, H., Zeilinger, A., Zukowski, M.: Multiphoton entanglement and interferometry. Rev. Mod. Phys. \textbf{84}, 777 (2012)


\bibitem{CGHZ0} Fr\"{o}wis F. and D\"{u}r W.: Stable macroscopic quantum superpositions. Phys. Rev. Lett. \textbf{106}, 110402 (2011)

\bibitem{CGHZ3} Fr\"{o}wis F. and D\"{u}r W.: Stability of encoded macroscopic quantum superpositions. Phys. Rev. A \textbf{85}, 052329 (2012)



\bibitem{construct1} Ding, D., Yan, F. L.  Gao, T.: Preparation of km-photon concatenated Greenberger-Horne-Zeilinger states for observing distinctive quantum effects at macroscopic scales. J. Opt. Soc. Am. B \textbf{30}, 3075 (2013)

\bibitem{cghzanalysis1}Sheng, Y. B.,  Zhou, L.: Entanglement analysis for macroscopic Schr\"{o}inger Cat state. EPL \textbf{109}, 40009 (2015)

\bibitem{cghzanalysis2}Sheng, Y. B.,   Zhou, L.: Two-step complete polarization logic Bell-state analysis. Sci. Rep. \textbf{5}, 13453 (2015)

\bibitem{cghzanalysis3} Zhou, L., Sheng, Y. B.: Complete logic Bell-state analysis assisted with photonic Faraday rotation. Phys. Rev. A \textbf{92}, 042314 (2015)


\bibitem{cghzanalysis4} Zhou, L.,  Sheng, Y. B.: Feasible logic Bell-state analysis with linear optics. Sci. Rep. \textbf{6}, 20901 (2016)

\bibitem{cghzpurification}Zhou, L.,  Sheng, Y. B.: Purification of logic-qubit
entanglement. Sci. Rep. \textbf{6}, 28813 (2016)

\bibitem{cghzconcentration1} Qu, C. C., Zhou, L., and Sheng, Y. B.: Entanglement concentration for concatenated Greenberger-Horne-Zeilinger state. Quant. Inf.
Process. \textbf{14}, 4131-4146 (2015)

\bibitem{cghzconcentration2} Pan, J., Zhou, L., Gu, S. P., Wang, X., F., Sheng, Y. B., Wang, Q.: Efficient entanglement concentration for concatenated Greenberger-Horne-Zeilinger state with the cross-Kerr
nonlinearity. Quant. Inf. Process. \textbf{15}, 1669-1687 (2016)


\bibitem{construct2} Lu, H., Chen, L.K., Liu, C., Xu, P., Yao, X. C., Li, L., Liu, N.L., Zhao, B., Chen, Y.A.,  Pan, J. W.: Experimental realization of a concatenated Greenberger-Horne-Zeilinger state for macroscopic quantum superpositions. Nat. Photon. \textbf{8}, 364-368 (2014)


\bibitem{polarization}Knill, E., Laflamme, R. and Milburn, G.J.: A scheme for efficient quantum computation with linear optics. Nature \textbf{409}, 46 (2001)

\bibitem{squeezed0}Braunstein, S. L., Kimble, H. J.:Teleportation of continuous quantum variables.  Phys. Rev. Lett. \textbf{80}, 869 (1998)

\bibitem{squeezed1}Zhang, Y. C., Li, Z. Y., Yu, S., Gu, W. Y., Peng, X. Guo, H.:Continuous-variable measurement-device-independent quantum key distribution using squeezed states. Phys. Rev. A \textbf{90}, 052325 (2014)

\bibitem{squeezed2}Hao, S. H., Su, X. L., Tian, C. X., Xie, C. D., Peng, K. C.:Five-wave-packet quantum error correction based on continuous-variable cluster entanglement. Sci. Rep. \textbf{5}, 15462 (2015)

\bibitem{squeezed3}Deng, X. W., Hao, S. H., Guo, H., Xie, C. D., Su, X. L.:
Continuous variable quantum optical simulation for time evolution of quantum harmonic oscillators. Sci. Rep. \textbf{6}, 22914 (2016)

\bibitem{squeezed4}Deng, X. W., Hao, S. H., Tian, C. X., Su, X. L., Xie, C. D., Peng, K. C.:
Disappearance and revival of squeezing in quantum communication with squeezed state over a noisy channel. Appl. Phys. Lett. \textbf{108}, 081105 (2016)
\bibitem{Gaussian0}Su, X. L.: Applying Gaussian quantum discord to quantum key distribution.  Chin. Sci. Bull. \textbf{59}, 1083 (2014)

\bibitem{Gaussian1}Zhang, Y. C., Yu, S., Guo, H.:Application of practical noiseless linear amplifier in no-switching continuous-variable quantum cryptography. Quant. Inf. Process. \textbf{14}, 4339-4349 (2015)

\bibitem{Gaussian2} de Faria, A. J.: Nondestructive verification of continuous-variable entanglement. Phys. Rev. A \textbf{94}, 012301 (2016)




\bibitem{ECS0}  Sanders, B. C.: Entangled coherent states. Phys. Rev. A \textbf{45}, 6811 (1992)

\bibitem{ECS1}Wang, X. G.:Quantum teleportation of entangled coherent states.  Phys. Rev. A \textbf{64}, 022302 (2001)



\bibitem{ECS2}  Jeong, H. Kim, M. S.:Efficient quantum computation using coherent states. Phys. Rev. A \textbf{65}, 042305 (2002)


\bibitem{ECS3} Jeong, H.  An, N. B.: Greenberger-Horne-Zeilinger-type and W-type entangled coherent states: Generation and Bell-type inequality tests without photon counting. Phys. Rev. A \textbf{74}, 022104 (2006)

\bibitem{ECSadd2}Park, C. Y., Jeong, H.:Bell-inequality tests using asymmetric entangled coherent states in asymmetric lossy environments.  Phys. Rev. A \textbf{91}, 042328 (2015)

\bibitem{ECS4}An, N. B.:Teleportation of coherent-state superpositions within a network. Phys. Rev. A \textbf{68}, 022321 (2003)

\bibitem{ECS5}An, N. B.: Optimal processing of quantum information via W-type entangled coherent states. Phys. Rev. A \textbf{69}, 022315 (2004)


\bibitem{ECS6}Sheng, Y. B., Liu, J., Zhao, S. Y., Wang, L., Zhou, L.:Entanglement concentration for W-type entangled coherent states. Chin. Phys. B \textbf{23}, 080305 (2014)

\bibitem{ECS7}Sheng, Y. B., Qu, C. C., Ou-Yang, Y., Feng, Z. F., Zhou L:Practical entanglement concentration for entangled coherent states. Int. J. Theor. Phys.  \textbf{53}, 2033-2040 (2014)

\bibitem{ECSadd1}Li, Z. Y., Zhang, Y. C., Wang, X. Y., Xu, B. J., Peng, X., Guo, H.:Non-Gaussian postselection and virtual photon subtraction in continuous-variable quantum key distribution. Phys. Rev. A \textbf{93}, 012310 (2016)

\bibitem{ECSadd3}Guo, R., Zhou, L., Gu, S. P., Wang, X. F., Sheng, Y. B.:Hybrid entanglement concentration assisted with single coherent state. Chin. Phys. B \textbf{25}, 030302 (2016)

\bibitem{ECSadd4}Jeong, H., Bae, S., Choi, S.:Quantum teleportation between a single-rail single-photon qubit
 and a coherent-state qubit using hybrid entanglement under decoherence effects. Quant. Inf. Process. \textbf{15}, 913-927 (2016)

 \bibitem{ECSadd5}Wei, C. P., Hu, X. Y., Yu, Y. F., Zhang, Z. M.:Phase sensitivity of two nonlinear interferometers with inputting entangled coherent states. Chin. Phys. B \textbf{25}, 040601 (2016)
\bibitem{ECS8}Tipsmark, A., Dong, R., Laghaout, A., Marek, P., Jezke, M. Andersen, U. L.:Experimental demonstration of a Hadamard gate for coherent state qubits. Phys. Rev. A \textbf{84}, 050301 (2011)





\end{thebibliography}
\end{document}